\documentclass[sigconf]{acmart}

\AtBeginDocument{%
  \providecommand\BibTeX{{%
    \normalfont B\kern-0.5em{\scshape i\kern-0.25em b}\kern-0.8em\TeX}}}

\setcopyright{acmcopyright}
\copyrightyear{2018}
\acmYear{2018}
\acmDOI{10.1145/1122445.1122456}

\acmConference[Woodstock '18]{Woodstock '18: ACM Symposium on Neural
  Gaze Detection}{June 03--05, 2018}{Woodstock, NY}
\acmBooktitle{Woodstock '18: ACM Symposium on Neural Gaze Detection,
  June 03--05, 2018, Woodstock, NY}
\acmPrice{15.00}
\acmISBN{978-1-4503-XXXX-X/18/06}

\begin{document}

\title{audino: A Modern Annotation Tool for Audio and Speech}

\author{Manraj Singh Grover}
\affiliation{%
  \institution{Indraprastha Institute of Information Technology}
  \state{Delhi}
  \country{India}
}
\email{manrajg@iiitd.ac.in}

\author{Pakhi Bamdev}
\affiliation{%
  \institution{Indraprastha Institute of Information Technology}
  \state{Delhi}
  \country{India}}
\email{pakhii@iiitd.ac.in}

\author{Ratin Kumar Brala}
\affiliation{%
  \institution{National Institute of Technology, Kurukshetra}
  \state{Haryana}
  \country{India}}
\email{ratin\_11822004@nitkkr.ac.in}

\author{Yaman Kumar}
\affiliation{%
  \institution{Indraprastha Institute of Information Technology}
  \state{Delhi}
  \country{India}}
\email{yamank@iiitd.ac.in}

\author{Mika Hama}
\affiliation{%
  \institution{Second Language Testing Inc.}
  \state{Princeton}
  \country{United States}}
\email{mika.hama@2lti.com}

\author{Rajiv Ratn Shah}
\affiliation{%
  \institution{Indraprastha Institute of Information Technology}
  \state{Delhi}
  \country{India}}
\email{rajivratn@iiitd.ac.in}


\begin{abstract}
  In this paper, we introduce a collaborative and modern annotation tool for audio and speech: audino. The tool allows annotators to define and describe temporal segmentation in audios. These segments can be labelled and transcribed easily using a dynamically generated form. An admin can centrally control user roles and project assignment through the admin dashboard. The dashboard also enables describing labels and their values. The annotations can easily be exported in JSON format for further analysis. The tool allows audio data and their corresponding annotations to be uploaded and assigned to a user through a key-based API. The flexibility available in the annotation tool enables annotation for Speech Scoring, Voice Activity Detection (VAD), Speaker Diarisation, Speaker Identification, Speech Recognition, Emotion Recognition tasks and more. The MIT open source license allows it to be used for academic and commercial projects.
\end{abstract}



\maketitle

\section{Introduction}
\label{sec:intro}


Over the past few years, there has been a dramatic improvement in audio and speech research. The rise and performance of deep neural network have achieved state-of-the-art results on various speech and audio tasks \cite{Li2019,8639583,Fujita2019Interspeech,grover2020multimodal}. These networks are necessary to consume and discover information in large volumes of data being published on the web. This necessitates the need to annotate data efficiently at scale for supervised learning of network. In this paper, we present a flexible and modern web-based annotation tool for audio and speech data called audino. The tool aims to provide a broad set of features required for annotation of speech datasets while focusing on increasing collaboration, project management and accessibility. The annotation tool is permissively licensed MIT\footnote{https://opensource.org/licenses/MIT} allowing it to be freely used for both academic research as well as commercial use. The tool can be downloaded from \texttt{https://github.com/midas-research/audino}.

Many annotation tools already exist for image~\cite{ 10.1145/3240508.3243656, labelme2016}, text~\cite{brat2012,eckart-de-castilho-etal-2016-web,doccano} and speech~\cite{BARRAS20015,glenn2009xtrans,kipp2001anvil} modality, where most of them require software installation on annotator's system. Recently, there has been increased interest in developing web-based annotation tools~\cite{brat2012,doccano,winkelmann-raess-2014-introducing,10.1145/3343031.3350535,Label_Studio,Levy2019}. Moving annotation tools to the web offer several advantages including data security, management and accessibility. A large number of these tools allow loading data, processing and saving annotations on annotator's web browser, while others offer server-side data loading and annotation storage. For speech modality, however, none of the open-source annotation tools to our knowledge offered complete management of annotation process and other advantages of a server-side annotation tool at the time of development of the project (see Table~\ref{tab:comparison} for comparsion between various web-based audio annotation tools available). With this motivation, we developed audino.

\begin{table*}[]
    \centering
    \begin{tabular}{|*{6}{c|}}
        \hline
        \textbf{Features} & \textbf{VIA \cite{10.1145/3343031.3350535}} & \textbf{gecko \cite{Levy2019}} & \textbf{Label Studio \cite{Label_Studio}} & \textbf{EMU-webApp \cite{winkelmann-raess-2014-introducing}} & \textbf{audino} \\
        \hline
        Segmentation and Labelling & \checkmark & \checkmark & \checkmark & \checkmark & \checkmark \\
        Project Management & \texttimes & \texttimes & \texttimes & \texttimes & \checkmark \\
        User Management & \texttimes & \texttimes & \texttimes & \texttimes & \checkmark \\
        Label Management & \texttimes & \texttimes & \checkmark & \checkmark & \checkmark \\
        Cloud Storage Access & \texttimes & \checkmark & \checkmark & \texttimes & \texttimes \\
        Database Storage & \texttimes & \texttimes & \texttimes & \checkmark & \checkmark \\
        Server side & \texttimes & \texttimes & \checkmark & \checkmark & \checkmark \\
        License & BSD 2 & BSD 3 & Apache 2.0 &MIT& MIT \\
        \hline
    \end{tabular}
    \caption{Comparison between various open source web-based audio annotation tools.}
    \label{tab:comparison}
\end{table*}

We share and discuss the salient features of the tool below:

\begin{itemize}
    \item \textbf{Accessibility.} In contrast to most annotation tools which need to be installed and run on annotator's system, audino is a web-based tool which makes it much easier to access through a web browser remotely. The side-effect of having data on annotator's system and the need to load a new datapoint after every datapoint annotation completion is mitigated.
    \item \textbf{Centralized control of data allocation, project management, and annotations.} In contrast to offline tools available, audino secures data access and simplifies project management through centralization. All labels are controlled centrally, which makes it less prone to errors. The annotations are saved in a central database, making it easier to consume.
    \item \textbf{Easy setup and deployment.} The project makes use of Docker to deliver the software easing the setup process, deployment and also scaling of the tool.
    \item \textbf{Security.} The application implements JSON Web Token~\cite{Jones2015} based authentication and authorization for secure login. An annotator can only view projects they are part of and can only access datapoints assigned to them.  The audio filenames of all data points are hashed to prevent remote scraping further increases data security.
    \item \textbf{Multi-language and emoji support.} The tool supports Unicode character set, which enables annotation of multi-language datasets for tasks like Code-Switched~\cite{auer2013code} Automated Speech Recognition (ASR).
\end{itemize}

\begin{figure}[t]
  \centering
  \includegraphics[width=\linewidth]{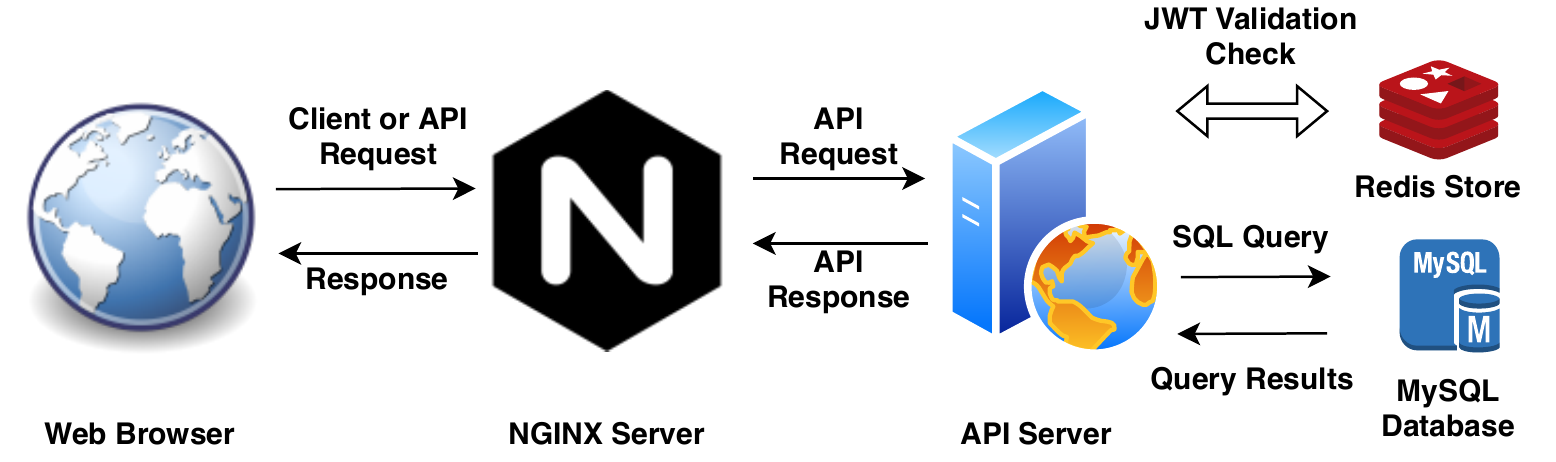}
  \caption{High level architecture of audino.}
  \label{fig:architecture}
\end{figure}

\section{Software Design}
\label{sec:softwaredesign}

audino is a production-ready web application tool. Figure~\ref{fig:architecture} provides a high-level overview of the working of different components in the tool. Its client-side is platform-independent and can run on any modern browser. The server side serves the REST API and static content. All annotations and application data are stored on the server.
We describe the software design in detail in the following sections.

\subsection{Data Storage}

The tool requires three types of data to be stored:

\subsubsection{Application data}
This includes users, roles, projects, data, labels, and annotations generated. This data is stored in a structured format in a dockerized SQL database. The entity-relationship diagram for the database is shared in documentation\footnote{https://github.com/midas-research/audino/tree/master/docs/database}. For the current version, the tool supports MySQL database, however, it can easily be extended to other SQL databases available.

\subsubsection{User session} To store the current user session, the application uses a dockerized Redis\footnote{https://redis.io/} store. The application generates JSON Web Token ID for every user login and saves it in the store (with an expiration time) for future authentication.

\subsubsection{Audio data} The audio uploaded is saved at a defined path inside the backend docker container. The application generates a unique filename for each uploaded file and stores the name inside the SQL database. The application then serves this file on request. The tool currently supports WAV, MP3 and OGG file formats as all browsers widely support these.


\subsection{Server Side}

In addition to storage components discussed in the previous section, the server-side of the tool also includes an NGINX\footnote{https://www.nginx.com/} server and an API server. NGINX is a high-performance web server which can also be used for reverse proxy and HTTP cache. The application utilizes a dockerized NGINX server to serve static client-side content and a REST API using reverse proxy. The REST API server runs in a separate docker container using uWSGI\footnote{https://uwsgi-docs.readthedocs.io/}. The tool uses a Python-based framework called Flask\footnote{https://flask.palletsprojects.com/}, and its plugins to provide a RESTful API. This API allows authentication by checking the Redis store for request user's session. The API also enables the client-side to perform CRUD operations on the database. To interact and perform database operations, the API uses SQLAlchemy\footnote{https://www.sqlalchemy.org/} library. Also, it provides a layer over database allowing easy switching to other SQL databases available. Alembic\footnote{https://alembic.sqlalchemy.org/} library is used for versioning and migrating the database.

\subsection{Client Side}

The client-side interface is written mainly in HTML, CSS and JavaScript. The user interface is broken into individual components and developed using React\footnote{https://reactjs.org/}, a JavaScript library for building user interfaces. React allows wiring of these client-side components with respective handlers as well as the REST API. Based on user interactions, the React components are rendered, and API requests are made. To make the application work for all screen size, the interface is styled using Bootstrap\footnote{https://getbootstrap.com/} CSS framework. The annotation dashboard leverages wavesurfer.js\footnote{https://wavesurfer-js.org/} library and its plugins for rendering audios and marking temporal regions. A production build is generated using React build system for NGINX to serve.

\section{Workflow}
\label{sec:workflow}

\begin{figure}[t]
  \centering
  \includegraphics[width=\linewidth]{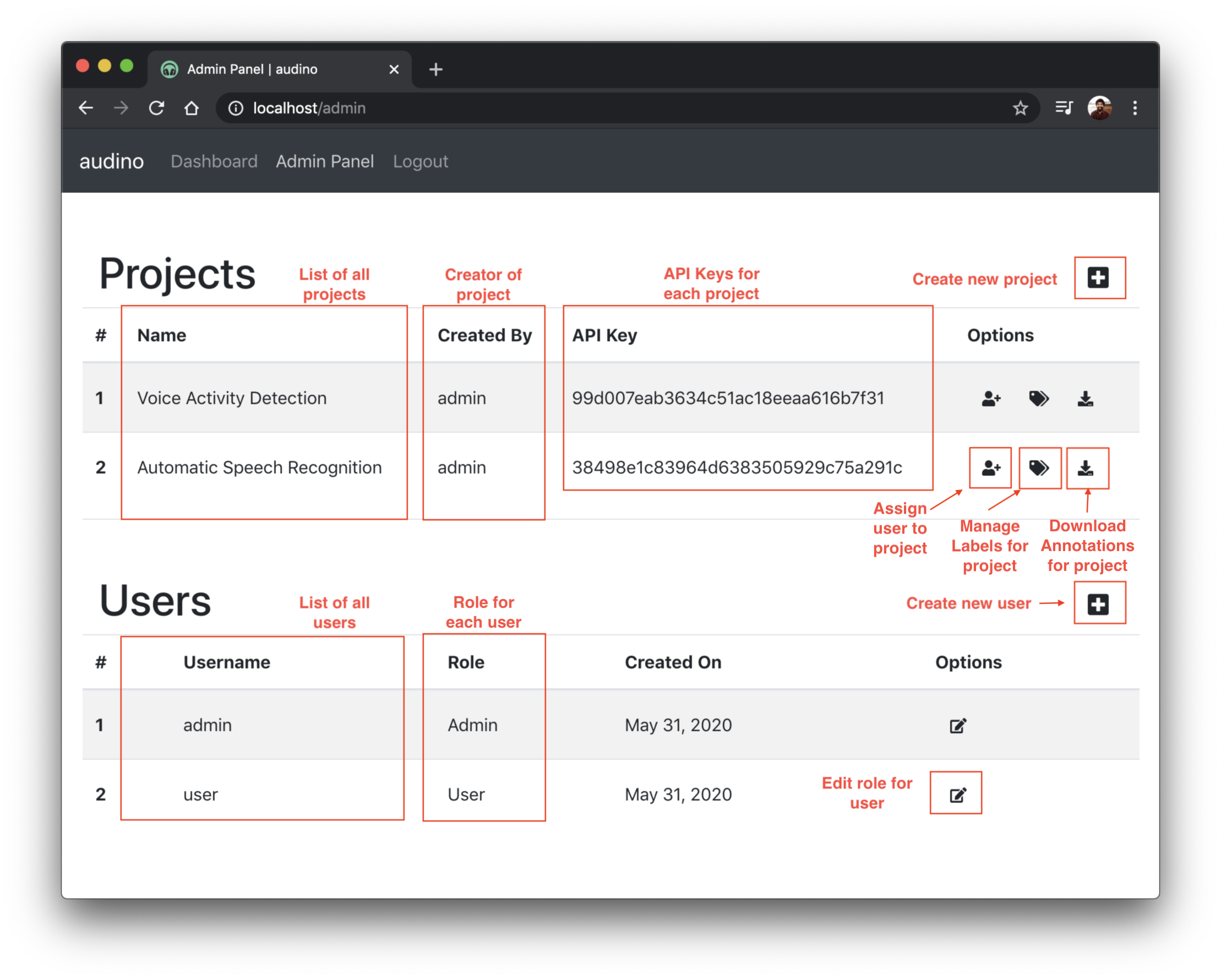}
  \caption{Admin panel with marked regions
explaining various functionalities available and accessible to users with admin roles.}
  \label{fig:admindashboard}
\end{figure}

An admin account is created during setup based on the information provided by the user. Account details of this user should be used to login for the first time. On opening the web application, a login screen is displayed. A user is required to have their account details in order to access the tool. Once logged in, the user dashboard is displayed listing the projects assigned to the user. The user can click on a project name to move to a dashboard which lists audio datapoints assigned to that user for that project in a paginated manner. These datapoints are categorized based on their completion status and whether they are marked for review or not. On clicking on the filename of a datapoint, the annotation panel opens for that audio. We will describe the annotation panel in detail in Section~\ref{sec:annotationdashboard}.

The application also provides an admin panel accessible to users with admin role (illustrated in Figure~\ref{fig:admindashboard}). This panel allows admins to manage projects and users. An admin can create new users, assign roles and projects to that user through this panel. The panel also allows the creation of new projects, labels and their associated label values for that project. These labels can be single choice select or multi-choice depending on the requirement. For each new project, an API Key is generated, which allows uploading of new datapoints and, optionally, their corresponding reference transcriptions and annotations for that project. This gives flexibility to pre-generate noisy labels or transcription for specific tasks using pre-trained machine learning models, review and correct them using the tool. Post the completion of annotation process, the annotations can be exported through the panel and consumed further for analysis or in any machine learning pipeline.

\section{Annotation Dashboard}
\label{sec:annotationdashboard}

\begin{figure*}[tb]
  \centering
  \includegraphics[width=0.57\linewidth]{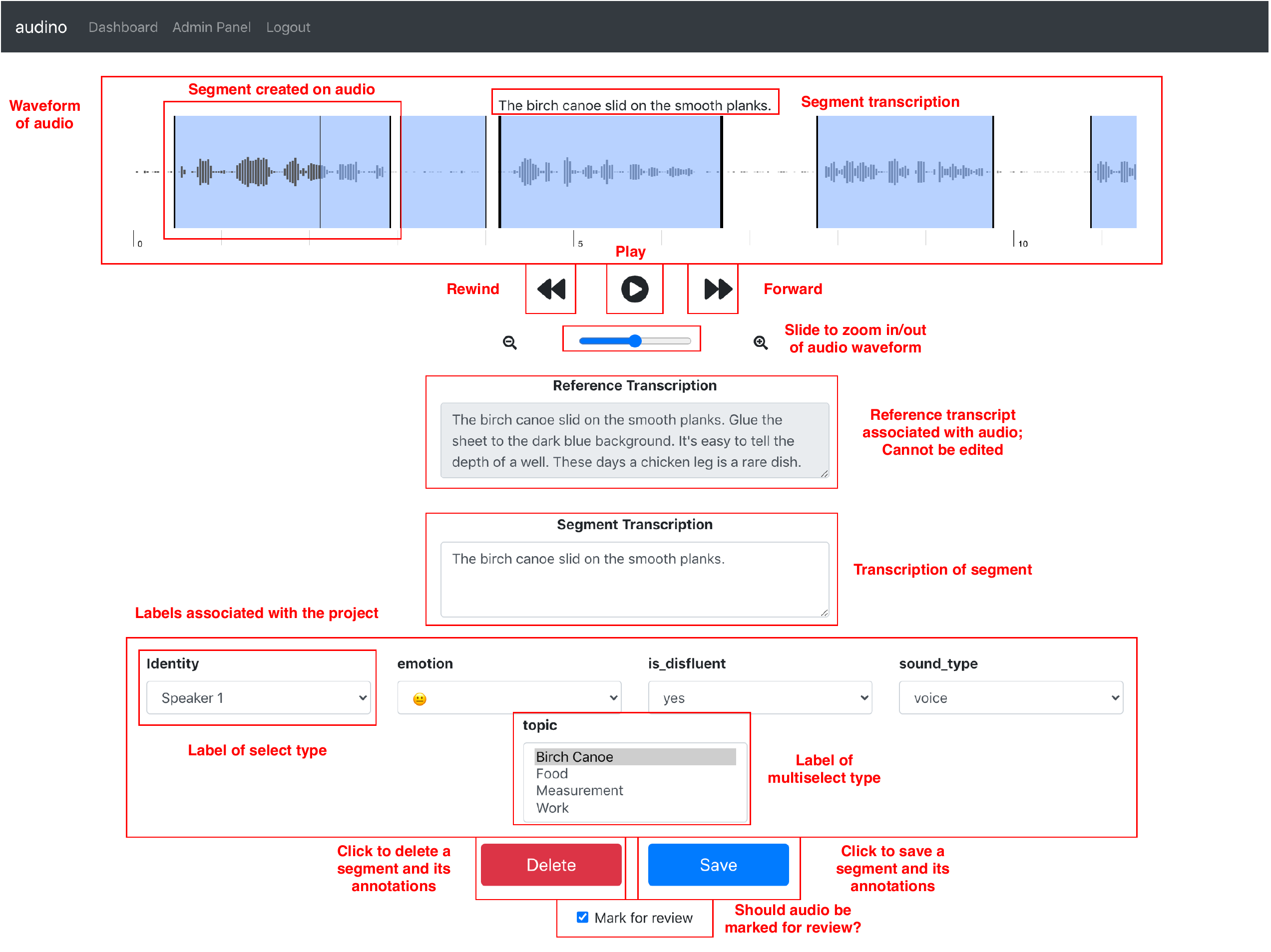}
  \caption{Screenshot of annotation dashboard showcasing various components.}
  \label{fig:annotationdashboard}
\end{figure*}

Figure~\ref{fig:annotationdashboard} illustrates the annotation dashboard. The audio datapoint selected is rendered as a waveform. This component allows users to create multiple temporal segments on audio for annotation. An audio control panel is provided to pause/play, move forward and backwards on the audio timeline. A zoom slider is also provided to control and zoom into a particular audio section for precise segmentation. This is particularly useful for annotating segments of phonemes. A reference transcription is displayed below the control panel, if provided when the datapoint was uploaded. These reference transcriptions could be ASR generated or transcribed by humans. This is specifically useful for tasks like speaker diarisation and ASR, where efficient segmentation and correction of transcriptions is required for improving ASR performance. On segment selection, a form consisting of segment transcript and associated project labels is displayed and is to be filled by the annotator. The annotator can save or delete any segment during the process, and the same will be reflected in the database. Finally, the annotators can mark a datapoint for review. These datapoints are displayed under a separate category on the project's data dashboard.

\section{Use Cases}
\label{sec:casestudy}

A primary use of audino is for transcription of human speech. We successfully used audino to transcribe the data collected through the administration of Simulated Oral Proficiency Interview for L2 English speakers by Second Language Testing, Inc., a US-based language testing company, to screen potential employees. The audios were sampled and assigned to two transcribers with an overlap of 20\% for quality check. The transcribers segmented the audios and transcribe each segment on the tool. These transcriptions were consumed by a CRON job which tracked and reported Word Error Rate (WER) between the two transcribers. It also flagged the audios with major transcription errors which were later resolved through discussion. Using audino, a total of 65 hours of responses were transcribed which, later, was used to fine-tune a pre-trained ASR. This ASR system was later used for transcription of complete dataset and developing an automated oral proficiency scoring system \cite{grover2020multimodal}. Among other projects, the audino tool has been used for annotating Content, Coherence and Disfluency attributes for the same data.

\section{Summary and Roadmap}
\label{sec:summaryandroadmap}

In this paper, we presented audino, a collaborative web-based modern annotation tool that allows temporal segmentation, transcription and labelling of language and speech aspects. We provide comprehensive documentation and tutorials to get the users started on our GitHub page. The project has been under active development for more than a year now and has been used successfully for large-scale projects at our lab. Open sourcing the tool allows us to discover new possibilities of its utilization while enabling collaboration and easier management of dataset generation task.

The short-term roadmap of the project includes adding enhancements like providing connectors for usage of cloud storage, keyboard shortcuts, improve mobile experience, exposing complete API through API key authentication, and an analytics dashboard, which can offer insights into the quality of annotations generated, their statistics and agreement between annotators. Some of these features are already in development phase and will be released soon. The long-term roadmap includes improving test coverage of the project, adding continuous integration and delivery to development flow, adding project templates for speech-related tasks enabling more straightforward project setup, and leveraging recent state-of-the-art models for automatic labelling and transcription of audios (reducing overall annotation effort). We welcome everyone to contribute to the project and provide constructive feedback.

\bibliographystyle{ACM-Reference-Format}
\bibliography{sample-sigconf}

\end{document}